\documentclass[runningheads,a4paper]{llncs}

\usepackage{amssymb}
\usepackage{graphicx}
\usepackage{amsmath}
\usepackage{mathpartir}

\usepackage{mathtools}

\usepackage{verbatim} 

\usepackage{stmaryrd} 
\usepackage{subfig}

\usepackage{wrapfig}

\usepackage{array}
\usepackage{isabelle}
\usepackage{isabellesym}

\usepackage[hyphens]{url}
\urldef{\mailsa}\path|{bzhan,haslbema}@in.tum.de|    
\newcommand{\keywords}[1]{\par\addvspace\baselineskip
\noindent\keywordname\enspace\ignorespaces#1}

\newcommand{\arraymap}{\mapsto_a}

\newcommand{\isaTo}{\isasymRightarrow}

\newcommand{\isaImp}{\isasymLongrightarrow}

\newcommand{\polylog}{\operatorname{polylog}}
\newcommand{\polylogtwo}{\operatorname{polylog_2}}

\begin{document}

\mainmatter  

\title{Verifying Asymptotic Time Complexity of Imperative
Programs in Isabelle}

\titlerunning{Asymptotic Time Complexity in Isabelle}

%
%
\author{Bohua Zhan\and Maximilian P. L. Haslbeck%
}
\authorrunning{Bohua Zhan\and Maximilian P. L. Haslbeck} 

\institute{Technische Universit\"{a}t M\"{u}nchen\\
\mailsa\\
\url{http://www21.in.tum.de/~{zhan,haslbema}}}

%
%

\toctitle{Lecture Notes in Computer Science}
\tocauthor{Authors' Instructions}
\maketitle

\begin{abstract}
We present a framework in Isabelle for verifying asymptotic
time complexity of imperative programs. We build upon an extension
of Imperative HOL and its separation logic to include running time.
In addition to the basic arguments, our framework is able to
handle advanced techniques for time complexity analysis,
such as the use of the Akra--Bazzi theorem and amortized analysis.
Various automation is built and incorporated into the auto2
prover to reason about separation logic with time credits, and to
derive asymptotic behavior of functions. As case studies, we verify
the asymptotic time complexity (in addition to functional
correctness) of imperative algorithms and data structures such as
median of medians selection, Karatsuba's algorithm, and splay trees.

\end{abstract}


\section{Introduction}

In studies of formal verification of computer programs, most of
the focus has been on verifying functional correctness of a program.
However, for many algorithms, analysis of its running time can be as
difficult, or even more difficult than the proof of its functional
correctness. In such cases, it is of interest to verify the
run-time analysis, that is, showing that the algorithm, or a given
implementation of it, does have the claimed asymptotic
time complexity.

Interactive theorem provers are useful tools for performing such
a verification, as their soundness is based on a small
trusted kernel, hence long derivations can be made with a very
high level of confidence. So far, the work of Gu\'{e}neau et al.
\cite{fistful,chargueraud2017verifying}
appears to be the only general framework for asympotic 
time complexity analysis of imperative programs in an interactive
theorem prover. The framework is built in Coq, based on
Chargu\'{e}raud's CFML package \cite{Chargueraud_CFML} for verifying
imperative programs using characteristic formulas.

We present a new framework\footnote{available online at \url{https://github.com/bzhan/Imperative_HOL_Time}} for asymptotic
time complexity analysis in Isabelle/HOL \cite{Isabelle}.
The framework is an extension of Imperative HOL \cite{imphol}, which
represents imperative programs as monads. Compared
to \cite{fistful}, we go further in
two directions. First, we incorporate the work of Eberl
\cite{akrabazzi} on the Akra--Bazzi theorem to analyze several
divide-and-conquer algorithms. Second, we extend the auto2 prover
\cite{zhan18a} to provide substantial automation in reasoning
about separation logic with time credits, as well as deriving
asymptotic behavior of functions.

We also make use of existing work by Nipkow \cite{Nipkow-ITP15}
on analysis of amortized complexity for functional programs.
Based on this work, we verify the amortized complexity of
imperative implementations of two data structures: skew heaps and
splay trees.

Throughout our work, we place a great deal of emphasis on modular
development of proofs. As the main theorems to be proved are
concerned with asymptotic complexity rather than explicit constants,
they do not depend on implementation details.
In addition, by using an ad-hoc refinement scheme similar to that
in \cite{zhan18a}, the analysis of an imperative program is
divided into clearly-separated parts: proof of functional
correctness, analysis of asymptotic behavior of runtime functions,
and reasoning about separation logic. Further separation of concerns
is used in amortized analysis.

In summary, the main contributions of this paper are as follows:

\renewcommand{\labelitemi}{$\bullet$}
\begin{itemize}

\item We extend Imperative HOL and its separation logic to reason
about running time of imperative programs (Section~\ref{sec:intro_imp}).

\item We introduce a way to organize the verification so that proofs can
be divided cleanly into orthogonal parts (Section \ref{sec:strategy}).

\item We extend the existing setup of the auto2 prover for
separation logic to also work with time credits. We also set up
various automation for proving asymptotic behavior of functions in
one or two variables (Section~\ref{sec:automation}).

\item We demonstrate the broad applicability of our framework with
several case studies (Section~\ref{sec:examples}), including those involving advanced techniques
for runtime analysis such as the use of Akra--Bazzi theorem
(for merge sort, median of medians selection, and Karatsuba's
algorithm) and amortized analysis (for dynamic arrays, skew heaps,
and splay trees). We also provide an example (Knapsack problem)
illustrating asymptotic complexity on two variables.

\end{itemize}
\renewcommand{\labelitemi}{--}


\section{Background}
\label{sec:background}

In this section, we review some background material needed in
our work. First, we briefly describe the extension of Imperative
HOL to reason about running time of imperative programs.
Then, we recapitulate the existing theory of asymptotic complexity
in Isabelle, and Eberl's formalization of the Akra--Bazzi theorem.

\subsection{Imperative HOL with time}
\label{sec:intro_imp}

Imperative HOL \cite{imphol} is a framework for
reasoning about imperative programs in Isabelle.
Lammich and Meis later constructed a separation logic 
for this framework \cite{afp-sep}. More details on both can be found in \cite{refinement}.

Atkey \cite{atkey2010amortised} introduced the idea of including
\emph{time credits} in separation logic to enable amortized 
resource analysis, in particular analysis of the running time of a
program. In this section, we describe how this idea is implemented
by modifying Imperative HOL and its separation logic.

\subsubsection{Basic definitions}

In ordinary Imperative HOL, a procedure takes a heap (of type
\isa{heap}) as input, and can either fail, or return a pair
consisting of a return value and a new heap. In Imperative HOL
with time, a procedure returns in addition a natural number
when it succeeds, specifying the number of computation steps
used. Hence, the type of a procedure with return type \isa{'a}
is given by:
\[
\isa{datatype 'a Heap = Heap "heap \isaTo\ ('a $\times$ heap
$\times$ nat) option"}
\] 

In the separation logic for ordinary Imperative HOL, a
\emph{partial heap} is defined to be a heap together with a subset
of addresses (type \isa{heap $\times$ nat set}). In our case,
a partial heap can also contain a number of time credits.
Hence, the new type for partial heaps is given by
\isa{pheap = (heap $\times$ nat set) $\times$ nat}.
  
An assertion (type \isa{assn}) is, as before, a mapping from
\isa{pheap} to \isa{bool} that does not depend on values of the
heap outside the address set. 
The notation $((h,as),n) \vDash P$ means the partial heap $((h,as),n)$ satisfies the
assertion $P$.
The basic assertions have the
same meaning as before, except they also require the partial heap
to contain zero time credits. In addition we define the assertion $\$(n)$, to
specify a partial heap with $n$ time credits and nothing else.

The \emph{separating conjunction} of two assertions is defined as
follows (differences from original definition are marked in bold):
\[ P * Q = \lambda((h, as)\mathbf{,n}).\, \exists u\;v\;\mathbf{n_1\;n_2}.\begin{cases}\,u \cup v = as \wedge u
\cap v = \emptyset\; \wedge \mathbf{ n_1 + n_2 = n}\; \wedge \\ ((h, u),\mathbf{n_1})\vDash P \wedge ((h, v),\mathbf{n_2})\vDash Q.\end{cases} \]
That is, time credits can be split in a separation conjunction
in the same way as sets of addresses on the heap. In particular $\$(n+m)=\$n * \$m$.

\subsubsection{Hoare triples} A Hoare triple \isa{<$P$> $c$ <$Q$>} 
is a predicate of type
\[
\isa{assn \isaTo\ 'a Heap \isaTo\ ('a \isaTo\ assn) \isaTo\ bool},
\]
defined as follows: \isa{<$P$> $c$ <$Q$>} holds if for any
partial heap $((h, as),n)$ satisfying $P$, the execution of $c$ on
$h$ is successful with new heap $h'$, return value $r$, and time
consumption $t$, such that $n\ge t$, and the new partial heap
$((h',as'),n-t)$ satisfies $Q(r)$, where $as'$ is $as$
together with the newly allocated addresses. With this definition
of a Hoare triple with time, the frame rule continues to hold.

Most basic commands (e.g. accessing or updating a reference, getting
the length of an array) are defined to take one unit of 
computation time. Commands that operate on an entire array, for
example initializing an array, or extracting an array into a
functional list, are defined to take $n+1$ units of computation
time, where $n$ is the length of the array. From this, we can prove
Hoare triples for the basic commands. Some examples are:

\begin{isabelle}
\ \ <p $\arraymap$ xs * \$1> Array.len xs
<$\lambda$r. p $\arraymap$ xs * $\uparrow$(r = length xs)>
\end{isabelle}

\begin{isabelle}
\ \ <\$(n + 1)> Array.new n x
<$\lambda$r. r $\arraymap$ replicate n x>
\end{isabelle}

We define the notation \isa{<$P$> $c$ <$Q$>$_t$} as a shorthand for
\isa{<$P$> $c$ <$Q$ * true>}. The assertion \isa{true} holds for
any partial heap, and in particular can include any number of time
credits. Hence, a Hoare triple of the form
\isa{<$P$*\$(n)> $c$ <$Q$>$_t$} implies that the procedure $c$
costs \emph{at most} $n$ time credits. We very often state Hoare triples
in this form, and so only prove upper bounds on the
computation time of the program.

\subsection{Asymptotic analysis}

Working with asymptotic complexity informally can be particularly
error-prone, especially when several variables are involved. Some
examples of fallacious reasoning are given in
\cite[Section 2]{fistful}. In an interactive theorem proving
environment, such problems can be avoided, since all notions are
defined precisely, and all steps of reasoning must be formally
justified.

For the definition of the big-$O$ notation, or more generally Landau
symbols, we use the formalization by Eberl \cite{afp-landau},
where they are defined in a general form in terms of filters,
and therefore work also in the case of multiple variables.

In our work, we are primarily interested in functions of type
\isa{nat\isaTo real} (for the single variable case) and
\isa{nat$\times$nat\isaTo real} (for the two variables case).
Given a function $g$ of one of these types, the Landau symbols
$O(g), \Omega(g)$ and $\Theta(g)$ are sets of functions of the same
type. In the single variable case, using the standard filter
(\isa{at\_top} for \emph{limit at positive infinity}), the
definitions are as follows:
\begin{align*}
f \in O(g) &\longleftrightarrow \exists c > 0.\ 
\exists N.\ \forall n\geq N.\ |f(n)| \leq c \cdot |g(n)| \\
f \in \Omega(g) &\longleftrightarrow \exists c > 0.\ 
\exists N.\ \forall n\geq N.\ |f(n)| \geq c \cdot |g(n)| \\
f \in \Theta(g) &\longleftrightarrow f \in O(g)
\land f \in \Omega(g)
\end{align*}
In the two variable case, we will use the product filter
\isa{at\_top $\times_F$ at\_top} throughout. Expanding the definitions,
the meaning of the Landau symbols are as expected:
\begin{align*}
f \in O_2(g) &\longleftrightarrow \exists c > 0.\ 
\exists N.\ \forall n,m\geq N.\ |f(n,m)| \leq c \cdot |g(n,m)| \\
f \in \Omega_2(g) &\longleftrightarrow \exists c > 0.\ 
\exists N.\ \forall n,m\geq N.\ |f(n,m)| \geq c \cdot |g(n,m)| \\
f \in \Theta_2(g) &\longleftrightarrow f \in O_2(g)
\land f \in \Omega_2(g)
\end{align*}

\subsection{Akra--Bazzi theorem}\label{sec:akrabazzi}

A well known technique for analyzing the asymptotic time complexity
of divide and conquer algorithms is the Master Theorem 
(see for example \cite[Chapter 4]{cormen2009introduction}).
The Akra--Bazzi
theorem is a generalization of the Master Theorem to a wider range
of recurrences. Eberl \cite{akrabazzi} formalized the Akra--Bazzi
theorem in Isabelle, and also wrote tactics for applying this
theorem in a semi-automatic manner. Notably, the automation is
able to deal with taking ceiling and floor in recursive calls,
an essential ingredient for actual applications but often ignored
in informal presentations of the Master theorem.

In this section, we state a slightly simpler version of the result
that is sufficient for our applications.
Let $f : \mathbb{N} \rightarrow \mathbb{R}$ be a non-negative
function defined recursively as follows:

\begin{equation}
f(x) = g(x) + \sum\limits_{i=1}^{k}a_i \cdot f(h_i(x))
\quad\quad \text{for all } x \geq x_0
\end{equation}
where $x_0 \in \mathbb{N}$, $g(x)\geq 0$ for all $x\geq x_0$,
$a_i \geq 0$ and each $h_i(x) \in \mathbb{N}$ is either
$\lceil b_i \cdot x \rceil$ or $\lfloor b_i \cdot x \rfloor$ with
$0 < b_i < 1$.

The parameters $a_i$ and $b_i$ determine a single characteristic
value $p$, defined as the solution to the equation
\begin{equation}
\sum_{i=1}^k  a_i\cdot b_i^p = 1
\end{equation}

Depending on the relation between the asymptotic behavior of $g$
and $\Theta(x^p)$, there are three main cases of the Akra--Bazzi
theorem:
\begin{description}
    \item {Bottom-heavy:} if $g \in O(x^q)$ for $q < p$
    and $f(x)>0$ for sufficiently large $x$, then
    $f \in \Theta(x^p)$.
    \item {Balanced:} if $g \in \Theta(x^p\ln^{a}x)$ with $a\ge 0$,
    then $f \in \Theta(x^p\ln^{a+1}x)$.
    \item {Top-heavy:} if $g \in \Theta(x^q)$ for $q > p$, then
    $f \in \Theta(x^q)$.
\end{description}

All three cases are demonstrated in our examples
(in Karatsuba's algorithm, merge sort, and median of medians
selection, respectively).


\section{Organization of proofs}
\label{sec:strategy}

In this section, we describe our strategy for organizing the
verification of an imperative program together with its
time complexity analysis. The strategy is designed to achieve the
following goals:

\begin{itemize}
    \item Proof of functional correctness of the algorithm should
    be separate from the analysis of memory layout and time
    credits using separation logic.
    \item Analysis of time complexity should be separate from
    proof of correctness.
    \item Time complexity analysis should work with asymptotic
    bounds $\Theta$ most of the time, rather than with explicit
    constants.
    \item Compositionality: verification of an algorithm should
    result in a small number of theorems, which can be used in the
    verification of a larger algorithm. The statement of these
    theorems should not depend on implementation details.
\end{itemize}

We first consider the general case and then describe the additional layer
of organization for proofs involving amortized analysis.

\subsection{General case}
\label{sec:strategy_general}

For a procedure with name \isa{f}, we define three Isabelle functions:
\begin{description}
    \item {\isa{f\_fun}:} The functional version of the procedure.
    \item {\isa{f\_impl}:} The imperative version of the procedure.
    \item {\isa{f\_time}:} The runtime function of the procedure. 
\end{description}

The definition of \isa{f\_time} should be stated in terms of
runtime functions of procedures called by \isa{f\_impl},
in a way parallel to the definition of \isa{f\_impl}. If
\isa{f\_impl} is defined by recursion, \isa{f\_time} should also
be defined by recursion in the corresponding manner. 

The theorems to be proved are:
\begin{enumerate}
    \item The functional program \isa{f\_fun} satisfies the desired
    correctness property.
    \item A Hoare triple stating that \isa{f\_impl} implements
    \isa{f\_fun} and runs within \isa{f\_time}.
    \item The running time \isa{f\_time} satisfies the desired
    asymptotic behavior.
    \item Combining 1 and 2, a Hoare triple stating that \isa{f\_impl}
    satisfies the desired correctness property, and runs within
    \isa{f\_time}.
\end{enumerate}

Here the proof of Theorem 2 is expected to be routine, since the three
definitions follow the same structure.
Theorem 3 should involve only analysis of asymptotic behavior of functions,
while Theorem 1 should involve only reasoning with functional
data structures. In the end, Theorems 3 and 4 present an interface
for external use, whose statements do not depend on details of
the implementation or of the proofs.

We illustrate this strategy on the final step of merge sort. The
definitions of the functional and imperative programs are shown
side by side below. Note that the former
is working with a functional list, while the latter is working with
an imperative array on the heap. \newline

\begin{minipage}{.55\textwidth}
\centering
\begin{isabelle}
merge\_sort\_fun xs = \isanewline
\ (let n = length xs in \isanewline
\ \ (if n $\le$ 1 then xs \isanewline
\ \ \ else \isanewline
\ \ \ \ let as = take (n div 2) xs; \isanewline
\ \ \ \ \ \ \ \ bs = drop (n div 2) xs; \isanewline
\ \ \ \ \ \ \ \ as' = merge\_sort\_fun as; \isanewline
\ \ \ \ \ \ \ \ bs' = merge\_sort\_fun bs; \isanewline
\ \ \ \ \ \ \ \ r = merge\_list as' bs' \isanewline
\ \ \ \ in r \isanewline
\ \ ) \isanewline
\ )
\end{isabelle}
\end{minipage}%
\begin{minipage}{.45\textwidth}
\centering
\begin{isabelle}
merge\_sort\_impl X = do \{ \isanewline
\ \ n $\leftarrow$ Array.len X;  \isanewline
\ \ if n $\le$ 1 then return ()  \isanewline
\ \ else do \{   \isanewline
\ \ \ \ A $\leftarrow$ atake (n div 2) X;  \isanewline
\ \ \ \ B $\leftarrow$ adrop (n div 2) X;  \isanewline
\ \ \ \ merge\_sort\_impl A;  \isanewline
\ \ \ \ merge\_sort\_impl B;  \isanewline
\ \ \ \ mergeinto (n div 2) \isanewline\ \ \ \ \ \ \ (n - n div 2) A B X  \isanewline
\ \ \} \isanewline
\} 
\end{isabelle}
\end{minipage}\newline

The runtime function of the procedure is defined as follows:
\begin{isabelle}
\ \ n $\le$ 1 $\Longrightarrow$ merge\_sort\_time n = 2
\isanewline
\ \ n $>$ 1 $\Longrightarrow$ merge\_sort\_time n = 2 + atake\_time n + adrop\_time n +  \isanewline
\ \ \ \ \ \ merge\_sort\_time (n div 2) + merge\_sort\_time (n - n div 2)  +
\isanewline
\ \ \ \ \ \ mergeinto\_time n
\end{isabelle}

The theorems to be proved are as follows. First, correctness
of the functional algorithm \isa{merge\_sort\_fun}:

\begin{isabelle}
\ \ merge\_sort\_fun xs = sort xs
\end{isabelle}
Second, a Hoare triple asserting the agreement of the three
definitions:
\begin{isabelle}
\ \ <p $\arraymap$ xs * \$(merge\_sort\_time (length xs))> \isanewline
\ \ \ merge\_sort\_impl p \isanewline
\ \ <$\lambda$\_. p $\arraymap$ merge\_sort\_fun xs>$_t$
\isanewline
\end{isabelle}
Third, the asymptotic time complexity of \isa{merge\_sort\_time}:

\begin{isabelle}
\ \ merge\_sort\_time $\in$
$\Theta$($\lambda$n. n * ln n)
\end{isabelle}
Finally, Theorems 1 and 2 are combined to prove the final Hoare
triple for external use, with \isa{merge\_sort\_fun xs} replaced
by \isa{sort xs}.


\subsection{Amortized analysis}
\label{sec:amortised}

In an amortized analysis, we fix some type of data structure and
consider a set of primitive operations on it. For simplicity,
we assume each operation has exactly one input and output
data structure (extension to the general case is straightforward).
A potential function $P$ is defined on instances of the data
structure. Each procedure $f$ is associated an actual runtime
$f_t$ and an amortized runtime $f_{at}$. They are required to satisfy
the following inequality: let $a$ be the input data structure of $f$ 
and let $b$ be its output data structure, then
\begin{equation} \label{eq:amortized}
f_{at} + P(a) \ge f_t + P(b).\footnote{In many presentations, the amortized runtime $f_{at}$ is simply
\emph{defined} to be $f_t + P(b) - P(a)$. Our approach is more flexible
in allowing $f_{at}$ to be defined by a simple formula, and isolating the
complexity to the proof of (\ref{eq:amortized}).}
\end{equation}

The proof of inequality (\ref{eq:amortized}) usually involves
arithmetic, and sometimes the correctness of the functional algorithm.
For skew heaps and splay trees, the analogous results are already
proved in \cite{Nipkow-ITP15}, and only slight modifications are
necessary to bring them into the right form for our use.

The organization of an amortized analysis in our framework is as
follows. We define two assertions: the \emph{raw} assertion
\isa{raw\_assn t a} stating that the address $a$ refers to the
functional data structure $t$, and the \emph{amortized} assertion,
defined as
\begin{isabelle}
\ \ amor\_assn t a = raw\_assn t a * \$($P$(t)),
\end{isabelle}
where $P$ is the potential function.

For each primitive procedure \isa{f}, we define \isa{f\_fun},
\isa{f\_impl}, and \isa{f\_time} as before, where \isa{f\_time} is the
\emph{actual} runtime. We further define a function \isa{f\_atime} to be
the proposed amortized runtime. The theorems to be proved are as follows
(compare to the list in Section \ref{sec:strategy}):

\begin{itemize}
    \item [1.] The functional program \isa{f\_fun} satisfies the desired
    correctness property.
    \item [2.] A Hoare triple using the amortized
    assertion stating that \isa{f\_impl} implements \isa{f\_fun}
    and runs within \isa{f\_atime}, which is a consequence of the following:
    \begin{itemize}
    \item [2a.] A Hoare triple using the raw assertion stating that
    \isa{f\_impl} implements \isa{f\_fun} and runs within \isa{f\_time}.
    \item [2b.] The inequality between amortized and actual runtime.
    \end{itemize}
    \item [3.] The \emph{amortized} runtime \isa{f\_atime} satisfies the
    desired asymptotic behavior.
    \item [4.] Combining 1 and 2, a Hoare triple starting that \isa{f\_impl}
    satisfies the desired correctness property, and runs within
    \isa{f\_atime}.
\end{itemize}

In the case of data structures (and unlike merge sort), it is useful
to state Theorem 4 in terms of yet another, abstract assertion,
which hides the concrete reference to the data structure. 
This follows the technique described in \cite[Section 5.3]{zhan18a}.
Theorems 3 and 4 are the final results for external use.

We now illustrate this strategy using splay trees as an example.
The raw assertion is called \isa{btree}. The basic operation in a splay
tree is the ``splay'' operation, from which insertion and lookup can
be easily defined. For this operation, the functions \isa{splay},
\isa{splay\_impl}, and \isa{splay\_time} are defined by recursion
in a structurally similar manner. Theorem 3a takes the form:
\begin{isabelle}
\ \ <btree t a * \$(splay\_time x t)> \isanewline
\ \ \ splay\_impl x a \isanewline
\ \ <btree (splay x t)>$_t$
\end{isabelle}
Let \isa{splay\_tree\_P} be the potential function on splay trees.
Then the amortized assertion is defined as:

\begin{isabelle}
\ \ splay\_tree t a = btree t a * \$(splay\_tree\_P t)
\end{isabelle}
The amortized runtime for splay has a relatively simple expression:

\begin{isabelle}
\ \ splay\_atime n = 15 * ($\lceil$3 * log 2 n$\rceil$ + 2)
\end{isabelle}
The difficult part is showing the inequality relating actual and
amortized runtime (Theorem 2b):

\begin{isabelle}
\ \ bst t $\isaImp$ splay\_time x t + splay\_tree\_P (splay x t) $\le$
\isanewline
\ \ \ \ \ \ \ \ \ \ \ \ splay\_tree\_P t + splay\_atime (size1 t),
\end{isabelle}
which follows from the corresponding lemma in \cite{Nipkow-ITP15}.
Note the requirement that $t$ is a binary search tree. Combining
2a and 2b, we get Theorem 2:

\begin{isabelle}
\ \ bst t $\isaImp$ \isanewline
\ \ <splay\_tree t a * \$(splay\_atime (size1 t))> \isanewline
\ \ \ splay\_impl x a \isanewline
\ \ <splay\_tree (splay x t)>$_t$
\end{isabelle}
The asymptotic bound on the amortized runtime (Theorem 3) is:

\begin{isabelle}
\ \ splay\_atime $\in$ $\Theta$($\lambda$x. ln x)
\end{isabelle}
The functional correctness of \isa{splay} (Theorem 1) states that
it maintains sortedness of the binary search tree and its set of
elements:
\begin{isabelle} 
\ \ bst t $\isaImp$ bst (splay a t),
\ \ set\_tree (splay a t) = set\_tree t
\end{isabelle}
The abstract assertion hides the concrete tree behind an
existential quantifier:

\begin{isabelle}
splay\_tree\_set S a = ($\exists_A$t. splay\_tree t a *
$\uparrow$(bst t) * $\uparrow$(set\_tree t = S))
\end{isabelle}
The final Hoare triple takes the form:
\begin{isabelle}
\ \ <splay\_tree\_set S a * \$(splay\_atime (card S + 1))> \isanewline
\ \ \ splay\_impl x a \isanewline
\ \ <splay\_tree\_set S>$_t$
\end{isabelle}


\section{Setup for automation}
\label{sec:automation}

In this section, we describe automation to handle
two of the steps mentioned in the previous section:
one working with separation logic (for Theorem 2), and the other proving asymptotic behavior of
functions (for Theorem 3).

\subsection{Separation logic with time credits}
\label{sec:automation_sep}

First, we discuss automation for reasoning about separation logic
with time credits. This is an extension of the setup discussed in
\cite{zhan18a} for reasoning about ordinary separation logic.
Here, we focus on the additional setup concerning time credits.

The basic step in the proof is as follows: given an
assertion \isa{$h_i$ $\vDash$ $P$ * $\$T$} on the current heap,
and a Hoare triple
\begin{isabelle}
\ \ <$P'$ * $\$T'$ * $\uparrow$($b$)> $c$ <$Q$>
\end{isabelle}
for the next command, apply the Hoare triple to derive the
successful execution of $c$, and some assertion on the next heap
$h_{i+1}$. In ordinary separation logic (without
$\$T$ and $\$T'$), this involves matching $P'$ with parts of $P$,
proving the pure assertions $b$, and then applying the frame rule.
In the current case, we additionally need to show that $T'\le T$,
so $\$T$ can be rewritten as $\$(T'+T'')=\$T'*\$T''$ (we
assume that matching $P'$ with parts of $P$ instantiates all
arbitrary variables in $T'$).

In general, proving this inequality can involve arbitrarily complex
arguments. However, due to the close correspondence in the
definitions of \isa{a\_time} and \isa{a\_impl}, the actual tasks
usually lie in a simple case, and we engineer the automation to
focus on this case. First, we normalize both $T$ and $T'$ into
polynomial form:
\begin{equation}
T = c_1 p_1 + \dots + c_m p_m,\quad T' = d_1 q_1 + \dots + d_n q_n,
\end{equation}
where each $c_i$ and $d_j$ are constants, and each $p_i$ and $q_j$
are non-constant terms or 1. Next, for each term $d_j q_j$ in $T'$,
we try to find some term $c_i p_i$ in $T$ such that $p_i$ equals
$q_j$ according to the known equalities, and $d_j \le c_i$. If
such a term is found, we subtract $d_j p_i$ from $T$. This procedure
is performed on $T$ in sequence (so $d_2 q_2$ is searched on
the remainder of $T$ after subtracting $d_1 q_1$, etc). If the
procedure succeeds with $T''$ remaining, then we have
$T = T' + T''$.

The above procedure suffices in most cases. For example, given the
parallel definitions of \isa{merge\_sort\_impl} and
\isa{merge\_sort\_time} in Section \ref{sec:strategy_general},
it is able to show that \isa{merge\_sort\_impl} runs in time
\isa{merge\_sort\_time}. However, in some special cases, more is
needed. The extra reasoning often takes the following form: if
$s$ is a term in the normalized form of $T$, and $s \ge t$ holds
for some $t$ (an inequality that must be derived during the proof),
then the term $s$ can be replaced by $t$ in $T$.

In general, we permit the user to provide hints of the form
\begin{isabelle}
\ \ @have "s $\ge_t$ t",
\end{isabelle}
where the operator $\cdot\ge_t\cdot$ is equivalent to
$\cdot\ge\cdot$, used only to remind auto2 that the fact is for
modification of time credit only. Given this instruction, auto2
attempts to prove $s\ge t$, and when it succeeds, it replaces the 
assertion $h_i \vDash P * \$T$ on the current heap with
$h_i \vDash P * \$T' * \isa{true}$,
where the new time credit $T'$ is the normalized form of $T-s+t$.
This technique is needed in case studies such as binary
search and median of medians selection (see the explanation for
the latter in Section \ref{sec:examples}).

\subsection{Asymptotic analysis}
\label{sec:automation_run}

The second part of the automation is for analysis of asymptotic
behavior of runtime functions. Eberl \cite{afp-landau} already
provides automation for Landau symbols in the single variable case. 
In addition to incorporating it into our framework, we add facilities
for dealing with function composition and the two variable case. 

Because side conditions for the Akra--Bazzi theorem are in
the $\Theta$ form, we mainly deal with $\Theta$ and $\Theta_2$, stating
the exact asymptotic behaviors of running time functions.
However, since running time functions themselves are very often only
upper bounds of the actual running times, we are essentially still
proving big-$O$ bounds on running times of programs.

In our case, the general problem is
as follows: given the definition of \isa{f\_time$(n)$} as a sum
of terms of the form \isa{g\_time$(s(n))$} (runtime of procedures
called by \isa{f\_impl}), simple terms like $4n$ or $1$, or
recursive calls to \isa{f\_time}, determine the asymptotic
behavior of \isa{f\_time}.

To begin with, we maintain a table of the asymptotic behavior of
previously defined runtime functions. The attribute
\isa{asym\_bound} adds a new theorem to this table. This table can
be looked-up by the name of the procedure.

We restrict ourselves to asymptotic bounds of the form
\[ \polylog(a,b)=(\lambda n.\ n^a (\ln{n})^b), \]
where $a$ and $b$ are natural numbers.
In the two variable case,
we work with asymptotic bounds of the form
\[ \polylogtwo(a,b,c,d) = (\lambda (m,n).\ \polylog(a,b)(m) \cdot
\polylog(c,d)(n)). \]

This suffices for our present purposes, and can be extended in the
future. Note that this restriction does not mean our framework
cannot handle other complexity classes, only that they will require
more manual proofs (or further setup of automation).

\subsubsection{Non-recursive case}

When the runtime function is non-recursive, the analysis proceeds
by determining the asymptotic behavior of each of its summands,
then combine them together using the absorption rule: if
$g_1\in O(g_2)$, then $\Theta(g_1+g_2)=\Theta(g_2)$. Here, we make
use of existing automation in \cite{afp-landau} for deciding
inclusion of big-$O$ classes of polylog functions.

In addition, we prove the following composition rule: if 
$u\in\Theta(\polylog(a,b))$, and $v\in\Theta(\lambda n.\ n)$, then
$u\circ v\in\Theta(\polylog(a,b))$. This allows us to handle terms
of the form \isa{g\_time$(s(n))$} where $s$ is linear. Composition
is in general quite subtle: the analogous rule does not hold if $f$ is
the exponential function\footnote{
https://math.stackexchange.com/questions/761006/big-o-and-function-composition}.

We implemented automation for applying these rules, so that goals
such as the following can be solved automatically: if
$f_1\in\Theta(\lambda n.\ n)$ and $f_2\in\Theta(\lambda n. \ln n)$, then
\[ (\lambda n.\ f_1(n + 1) + n\cdot f_2(2n) + 3n\cdot f_2(n\ \isa{div}\ 3)) \in
\Theta(\lambda n.\ n \ln n). \]

Analogous results are proved in the two variable case (note that
unlike in the single variable case, not all pairs of $\polylogtwo$ functions
are comparable. e.g. $O(m^2n+mn^2)$). For example, the following can be automatically solved:
if additionally $f_3\in\Theta(\lambda (m,n).\ mn)$ and $f_4\in\Theta(\lambda (m,n).\ m + n)$, then
\begin{align*}
(\lambda (m,n).\ f_1(n) + f_2(m) + mn + f_3(m\ \isa{div}\ 3,n+1)) &\in \Theta(\lambda (m,n).\ mn). \\
(\lambda (m,n).\ 1 + f_1(n) + f_2(m) + f_4(m+1,n+1)) &\in \Theta(\lambda (m,n).\ m + n).
\end{align*}

\subsubsection{Recursive case}

There are two main classes of results for analysis of
recursively-defined runtime functions: the Akra--Bazzi theorem and
results about linear recurrences. 
For both classes of results, applying the theorem reduces the analysis
of a recursive runtime function to the analysis of a non-recursive
function, which can be solved using automation described in the
previous part.

The Akra--Bazzi theorem is discussed in Section \ref{sec:akrabazzi}. 
Theorems about linear recurrences
allow us to reason about for-loops written as recursions. They include
the following: in the single variable case, if $f$ is defined
by induction as
\[ f(0) = c,\quad f(n+1) = f(n) + g(n), \]
where $g\in\Theta(\lambda n.\ n)$, then $f\in\Theta(\lambda n.\ n^2)$.

In the two variable case, if $f$ satisfies
\[ f(0,m) \le C,\quad f(n+1,m) = f(n,m) + g(m) \]
where $g\in\Theta(\lambda n.\ n)$, then
$f\in\Theta_2(\lambda(n,m).\ nm)$.

As an example, consider the problem of showing
\isa{$\Theta$($\lambda$n. n * ln n)}
complexity of \isa{merge\_sort\_time}, defined in Section
\ref{sec:strategy_general}. This applies the balanced case of the
Akra--Bazzi theorem. Using this theorem, the goal is reduced to:
\begin{isabelle}
\ \ ($\lambda$n. 2 + atake\_time n + adrop\_time n +
mergeinto\_time n)
$\in$ $\Theta$($\lambda$n. n)
\end{isabelle}
(the non-recursive calls run in linear time). This can be shown
automatically using the method described in the previous section,
given that the previously-defined runtime functions \isa{atake\_time},
\isa{adrop\_time}, and \isa{mergeinto\_time} have already been shown
to be linear.


\section{Case studies}
\label{sec:examples}

In this section, we present the main case studies verified using
our framework. The examples can be divided into three classes:
divide-and-conquer algorithms (using the Akra--Bazzi theorem), algorithms
that are essentially for-loops (using linear recurrences), and amortized
analysis.

The following table lists for each case study the ratio (\#Ratio) between the sum of number 
of steps for the proof of Hoare triples (\#Hoare) and reasoning
about runtime functions (\#Time) to the number of lines of the imperative program (\#Imp).
This ratio measures the overhead for verifying the imperative program with runtime
analysis. In particular this does \emph{not} include verifying the correctness of the
functional program.
In addition we list the total lines of code for each case study. This includes
proofs as well as statements of definitions and lemmas.
\begin{center}
\begin{tabular}{l|>{\centering}m{1.5cm}|>{\centering}m{1.5cm}|>{\centering}m{1.5cm}|>{\centering}m{1.5cm}| c}
  & \#Imp & \#Time & \#Hoare & Ratio &  \quad LOC\quad\  \\ \hline
Binary search & 11 & 10 & 14 & 2.18 & 82 \\
Merge sort & 38 & 11 & 12 & 0.61 & 121 \\
Karatsuba & 58 & 18 & 28 & 0.79 & 250 \\
Select & 51 & 41 & 31 & 1.41 & 447 \\
Insertion sort & 15 & 3 & 4 & 0.47 & 42 \\
Knapsack & 27 & 9 & 8 & 0.63 & 113 \\
Dynamic array & 55 & 19 & 37 & 1.02 & 424 \\
Skew heap & 25 & 38 & 21 & 2.36 & 257 \\
Splay tree & 120 & 51 & 37 & 0.73 & 447 
\end{tabular}
\end{center}
Using our automation the average overhead ratio is slightly over 1. 
On a dual-core laptop with 2GHz each, processing all the examples takes
around ten minutes.

Next we give details for some of the case studies.

\subsubsection{Karatsuba's algorithm}

The functional version of Karatsuba's algorithm for multiplying two
polynomials is verified in \cite{Berlekamp_Zassenhaus-AFP}.
To simplify matters, we further restrict us to the case where the two
polynomials are of the same degree.

The recursive equation is given by:
\begin{equation}
T(n) = 2\cdot T(\lceil n/2\rceil) + T(\lfloor n/2\rfloor) + g(n).
\end{equation}
Here $g(n)$ is the sum of the running times corresponding to non-recursive calls,
which can be automatically shown to be linear in $n$.
Then the Akra--Bazzi method gives the solution
$T(n)\in\Theta(n^{log_2 3})$ (bottom-heavy case).

\subsubsection{Median of medians selection}

Median of medians for quickselect is a worst-case linear-time algorithm
for selecting the $i$-th largest element of an unsorted array
\cite[Section 9.3]{cormen2009introduction}. In the first step of the
algorithm, it chooses an approximate median $p$ by dividing the array
into groups of 5 elements, finding the median of each group, and
finding the median of the medians by a recursive call. In the second step, $p$ is
used as a pivot to partition the array, and depending on $i$ and
the size of the partitions, a recursive call may be made to either the
section $x<p$ or the section $x>p$. This algorithm is particularly
interesting because its runtime satisfies a special recursive formula:
\begin{equation}
T(n) \le T(\lceil n/5\rceil) + T(\lceil 7n/10\rceil) + g(n),
\end{equation}
where $g(n)$ is linear in $n$. Akra--Bazzi theorem shows that
$T$ is linear (top-heavy case).

Eberl verified the correctness of the functional algorithm
\cite{Median_Of_Medians_Selection-AFP}. There is one special difficulty
in verifying the imperative algorithm: the length of the array in
the second recursive call is not known in advance, only that it
is bounded by $\lceil 7n/10\rceil$. Hence, we need to prove
monotonicity of $T$, as well as provide the hint
$T(\lceil 7n/10\rceil) \ge_t T(l)$ (where $l$ is the
length of the array in the recursive call) during the proof.

\subsubsection{Knapsack}

The dynamic programming algorithm solving the Knapsack problem
is used to test our ability to handle asymptotic complexity with two variables.
The time complexity of the algorithm is $\Theta_2(nW)$, where $n$
is the number of items, and $W$ is the capacity of the sack.
Correctness of the functional algorithm was proved by Simon Wimmer.

\subsubsection{Dynamic array}

Dynamic Arrays \cite[Section 17.4]{cormen2009introduction}
are one of the simpler amortized data structures.
We verify the version that doubles the size of the array
whenever it is full (without automatically shrinking the array).

\subsubsection{Skew heap and splay tree}

For these two examples, the bulk of the analysis (functional correctness
and justification of amortized runtime) is done in \cite{Nipkow-ITP15}.
Our work is primarily to define the imperative version of the algorithm
and verifying its agreement with the functional version. Some work
is also needed to transform the results in \cite{Nipkow-ITP15} into
the appropriate form, in particular rounding the real-valued potentials
and runtime functions into natural numbers required in our framework.


\section{Related work}
\label{sec:RelWork}

We compare our work with recent advances in verification of
runtime analysis of programs, starting from those based on interactive
theorem provers to the more automatic methods.


The most closely-related is the impressive work by Gu\'{e}neau et al.
\cite{fistful} for asymptotic time complexity analysis in Coq.
We now take a closer look at the similarities and differences:
\begin{itemize}
    
\item Gu\'{e}neau et al. give a structured overview
of different problems that arise when working informally with
asymptotic complexity in several variables, then solve this problem
by rigorously defining asymptotic domination (which is essentially
$f\in O(g)$) with filters and develop automation for reasoning about it.
We build on existing formalization of Landau symbols 
with filters in Isabelle \cite{afp-landau}. We focus on establishing
$\Theta$ bounds (i.e. upper \emph{and} lower bounds) and extend automation
to handle the two-variables case.
    
\item While they package up the functional correctness together
with the complexity claims into one predicate \isa{specO}, we
choose to have two separate theorems (the Hoare triple and the asymptotic bound).
    
\item While their automation assists in synthesizing recurrence
equations from programs, they leave their solution to the
human.
In contrast, we write the recurrence relation by hand, which can
be highly non-obvious (e.g. in the case of median of medians selection),
but focus on solving the recurrences for the asymptotic bounds
automatically (e.g. using the Akra--Bazzi theorem).

\item They present several examples including binary search,
the Bellman--Ford algorithm and union-find.
We provide several advanced examples including those involving
the Akra--Bazzi method. We also demonstrate that verification
of amortized analysis of functional programs \cite{Nipkow-ITP15}
can be converted to verification of imperative programs with
little additional effort.

\end{itemize}


Wang et al. \cite{wang2017timl} present TiML, a functional
programming language which can be annotated by invariants and
specifically also with time complexity annotations in types.
The type checker extracts verification conditions from these
programs, which are handled by an SMT solver. They also make the
observation that annotational burden can be lowered by not
providing a closed form for a time bound, but only specifying its
asymptotic behaviour. For recursive functions, the generated VCs
include a recurrence (e.g. $T(n-1) + 4n \leq T(n)$)
and one is left to show that there exists a solution for $T$ which
is additionally in some asymptotic bound, e.g. $O(n^2)$.
By employing an recurrence solver based on heuristic pattern
matching they make use of the Master Theorem in order to discharge
such VCs. In that manner they are able to verify the asymptotic
complexity of merge sort.
Additionally they can handle amortized complexity, giving Dynamic
Arrays and Functional Queues as examples.
Several parts of their work rely on non-verified components, including
the use of SMT solvers and the pattern matching for recurrence relations.
In contrast, our work is verified throughout by Isabelle's kernel.

On the other end of the scale we want to mention Automatic
Amortized Resource Analysis (AARA). Possibly the first example of
a resource analysis logic based on potentials is due to Hofmann
and Jost \cite{DBLP:conf/esop/HofmannJ06}. They pioneer the use of
potentials coded into the type system in order to automatically
extract bounds in the runtime of functional programs.
Hoffmann et al. successfully developed this idea further \cite{hoffmann2011multivariate,hoffmann2017towards}.
Carbonneaux et al. \cite{carbonneaux2015compositional,carbonneaux2017automated}
extend this work to imperative programs and automatically solve
extracted inequalities by efficient off-the-shelf LP-solvers.
While the potentials involved are restricted to a specific shape,
the analysis performs well and at the same time generates Coq proof
objects certifying their resulting bounds.




\section{Conclusion}
\label{sec:conclusion} 
 
In this paper, we presented a framework for verifying asymptotic
time complexity of imperative programs. This is done by extending
Imperative HOL and its separation logic with time credits. Through
the case studies, we demonstrated the ability of our framework
to handle complex examples, including those involving advanced
techniques of time complexity analysis, such as Akra--Bazzi theorem
and amortized analysis.

One major goal for the future is to extend Imperative HOL with
\emph{while} and \emph{for} loops, and add facilities for reasoning
about them (both functional correctness and time complexity).
Ultimately, we would like to build a single framework in which all
deterministic algorithms typically taught in undergraduate study (for
example, those contained in \cite{cormen2009introduction}) can be
verified in a straightforward manner.

The Refinement Framework by Lammich \cite{refinement} is a
framework for stepwise refinement from specifications via deterministic
algorithms to programs written in Imperative HOL.
It would certainly be interesting to investigate how to combine this stepwise
refinement scheme with runtime analysis.

\subsubsection*{Acknowledgments.}
This work is funded by DFG Grant NI 491/16-1.
We thank Manuel Eberl for his impressive formalization of the Akra--Bazzi method and the functional correctness of the selection algorihtm, and Simon Wimmer for the formalization of the DP solution for the Knapsack problem. 
We thank Manuel Eberl, Tobias Nipkow, and Simon Wimmer for valuable feedback during the project.
Finally, we thank Arma\"{e}l Gu\'{e}neau and his co-authors for their stimulating paper.

\bibliographystyle{splncs03}
\bibliography{paper}

\end{document}